\documentclass[10pt]{sigplanconf}

\usepackage{url}                  
\usepackage{listings}          
\usepackage{enumitem}      

\usepackage{amsmath}
\usepackage{algorithm}
\usepackage{algpseudocode}
\usepackage{graphicx}
\usepackage{color}
\usepackage{fancyvrb}
\usepackage{subcaption}
\usepackage{flushend}

\definecolor{pblue}{rgb}{0.13,0.13,1}
\definecolor{pgreen}{rgb}{0,0.5,0}
\definecolor{pred}{rgb}{0.9,0,0}
\definecolor{pgrey}{rgb}{0.46,0.45,0.48}

\newcommand{\cut}[1]{}

\begin{document}
\toappear{}

\title{Prioritized Garbage Collection: \protect\\Explicit GC Support for Software Caches}
\authorinfo{Diogenes Nunez}{Tufts University, USA}{dan@cs.tufts.edu}
\authorinfo{Samuel Z. Guyer}{Tufts University, USA}{sguyer@cs.tufts.edu}
\authorinfo{Emery D. Berger}{University of Massachusetts Amherst, USA}{emery@cs.umass.edu}

\lstset{language=Java,
  showspaces=false,
  showtabs=false,
  breaklines=true,
  showstringspaces=false,
  breakatwhitespace=true,
  commentstyle=\color{pgreen},
  keywordstyle=\color{pblue},
  stringstyle=\color{pred},
  basicstyle=\scriptsize\ttfamily,
  moredelim=[il][\textcolor{pgrey}]{$$},
  moredelim=[is][\textcolor{pgrey}]{\%\%}{\%\%},
  captionpos=b
}

\maketitle

\begin{abstract}
Programmers routinely trade space for time to increase performance, often in the
form of caching or memoization. In managed languages like Java or JavaScript,
however, this space-time tradeoff is complex. Using more space translates into
higher garbage collection costs, especially at the limit of available
memory. Existing runtime systems provide limited support for
space-sensitive algorithms, forcing programmers into difficult and often brittle
choices about provisioning.

This paper presents \emph{prioritized garbage collection}, a cooperative
programming language and runtime solution to this problem. Prioritized GC
provides an interface similar to soft references, called \emph{priority
references}, which identify objects that the collector can reclaim eagerly if
necessary. The key difference is an API for defining the policy that governs
when priority references are cleared and in what order.  Application code
specifies a priority value for each reference and a target memory bound. The
collector reclaims references, lowest priority first, until the total memory
footprint of the cache fits within the bound.  We use this API to implement a
space-aware least-recently-used (LRU) cache, called a \emph{Sache}, that is a
drop-in replacement for existing caches, such as Google's Guava library. The
garbage collector automatically grows and shrinks the Sache in response to
available memory and workload with minimal provisioning information from the
programmer. Using a Sache, it is almost impossible for an application to
experience a memory leak, memory pressure, or an out-of-memory crash caused by
software caching.

\end{abstract}



\section{Introduction}

\begin{figure}[!b]
  \includegraphics[width=\columnwidth]{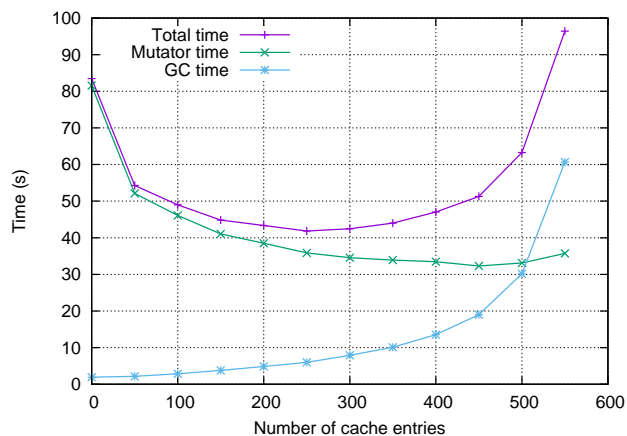}
  \caption{Competing tradeoffs: as cache size increases, miss time goes  
    down, but GC time goes up.}
  \label{fig:tradeoff}
\end{figure}

Software caching and garbage collection (GC) do not play well together. The
problem is that they embody conflicting goals and tradeoffs. Caching aims to
achieve the highest hit rate given a particular storage budget: the larger the
cache, the higher the hit rate. Unfortunately, most widely-used garbage
collection algorithms have a cost proportional to live memory~\cite{1094836}. In
this setting, the benefits of a larger cache are less clear because improvements in hit
rate are offset, to some degree, by additional GC costs. The penalty can become
particularly high if the cache starts competing with the rest of the program
for resources, increasing memory pressure and GC overhead.

Figure~\ref{fig:tradeoff} shows how the performance of a cache changes as we
vary the number of entries while keeping the overall heap size fixed. This
benchmark uses Google's Guava caching library configured with a
least-recently-used (LRU) eviction policy. We measured the time it takes
(Y-axis) to serve a predefined sequence of requests under each cache
configuration (X-axis). Looking from left to right, the two opposing trends are
clearly visible. Larger caches have higher hit rates, incurring fewer misses and
lowering mutator time. At the same time, though, larger caches increase GC
time. At the far right, the cache occupies almost all of available memory,
increasing not only the per-GC cost, but also the frequency of collection,
causing GC time to dominate total time. At the limit, the program runs out of
memory and crashes. Unbounded growth of caches (and related structures, such as
indexes) is a primary cause of memory leaks and performance problems in
Java~\cite{mitchell03leakbot, xu11leakchaser}.

The scenario above is a simple and controlled experiment -- real applications
have much more complex behavior, including multiple caches, possibly with
different eviction policies, different memory footprints, and different patterns
of locality, as well as significant non-cache data structures. Getting the most
out the available memory resources without triggering memory pressure poses a
significant challenge.

Existing runtime systems provide some mechanisms to support memory-sensitive
data structures, but they are sorely lacking in ways to configure and control
the policies that govern these mechanisms. As an example, the Java runtime
provides \emph{soft references}, which the garbage collector can clear at its
discretion to avoid running out of memory. A common programming strategy is to
store each cache entry in a soft reference, allowing the collector to reclaim
individual entries if necessary. The application, however, has little or no
control over when this process is triggered, or over which soft references are
cleared and in what order. Some Java Virtual Machines (JVMs), such as HotSpot,
use LRU-like policies that are clearly designed for caches, but are too coarse
for complex applications where a single global policy is not appropriate. In
Section~\ref{sec:problem} we describe this problem in more detail, and in
Section~\ref{sec:results} we show its effect on hit rate.

Not surprisingly, soft references are widely shunned. In fact, the official
documentation for the Android runtime library explicitly warns against using
them for caches~\cite{androidsoftrefs}: ``In practice, soft references are
inefficient for caching. The runtime doesn't have enough information on which
references to clear and which to keep. Most fatally, it doesn't know what to do
when given the choice between clearing a soft reference and growing the heap.
\emph{The lack of information on the value to your application of each reference
  limits the usefulness of soft references.} References that are cleared too
early cause unnecessary work; those that are cleared too late waste memory.''
In Section~\ref{sec:problem} we present detailed empirical measurements showing
that these are real problems.

\vspace{2em}

\noindent
This paper presents \emph{\bf prioritized garbage collection}, an automatic
memory management system designed to address the deficiencies outlined above by
providing explicit support for software caches and other space-sensitive data
structures. The key idea is to enable better cooperation between the application
and the garbage collector. In our system, the collector provides the
\emph{mechanisms} for measuring and enforcing memory usage, while the
application dictates the \emph{policies} that drive these mechanisms. The
application and the collector cooperate through a simple API that is designed
around a new kind of reference object we call a \texttt{PrioReference}
(short for ``priority reference''). It resembles a soft reference, except that
the application can specify both the \emph{global policy} (when to trigger
eviction and how much memory to reclaim) and the \emph{local policy} (which
priority references to clear and in what order). Our paper makes the following
contributions:

\begin{enumerate}

\item We quantify the performance problems of existing cache implementations by
  driving them with a range of workloads across a range of sizes. Not
  surprisingly, choosing a fixed cache size, particularly in terms of number of
  entries, is brittle. We also demonstrate the limitations of soft references as
  a mechanism for implementing these data structures.

\item We present a new reference type called \texttt{PrioReference} that allows
  application code to communicate the relative value of not reclaiming its
  referent object (and transitively, reachable objects). \texttt{PrioReference}s
  are grouped into \texttt{PrioSpace}s, which specify the details of the total
  memory limit and eviction policy for each group. A common configuration is one
  \texttt{PrioSpace} for each cache, but the mapping is up to the application to
  decide.

\item We describe the design and implementation of a garbage collector
  that enforces these policies. The key mechanism is a modified closure
  phase that visits \texttt{PrioReferences} in order from highest to lowest
  priority, stopping when the target space bound is reached. Unmarked
  references are implicitly evicted, and are reclaimed immediately by the
  sweeper without touching them.

\item We present a space-sensitive cache, which we call a \emph{Sache}, built on
  our new API. The Sache supports LRU and GreedyDual~\cite{cao97} eviction
  policies by changing the way it computes the reference priorities. Our system
  allows the user to set the target memory footprint in terms of available
  memory, so the Sache expands and contracts automatically to avoid memory
  pressure.

\item We report performance results obtained by driving a key-value store with a
  range of workloads. We use representative workloads to systematically explore
  the performance space and quantify the problems. We compare our cache to
  Google's Guava caching library on web traffic traces.\cut{ and on the Cache2k
  caching benchmark~\cite{some stuff}.}
\end{enumerate}

\noindent
The remainder of this paper is organized as
follows. Section~\ref{sec:problem} describes the problem in more detail and
explores the space of interactions between caches and garbage
collection. Section~\ref{sec:prioritized} describes the design and
implementation of our garbage collector mechanisms and the Sache data
structure. Section~\ref{sec:results} presents Sache performance results and
compares them to traditional caches. Finally, Sections~\ref{sec:related}
and~\ref{sec:conclusion} review related work and conclude.

\section{Problem}
\label{sec:problem}

It is not easy to implement software caching in a garbage collected language.
One reason is that cache performance is governed by a space-time tradeoff that
is in direct opposition to the tradeoff in garbage collection.  Another reason
is that garbage collected languages provide poor support for implementing
\emph{any} algorithm or data structure that is inherently space sensitive. In
this section, we discuss these issues in detail, and present measurements that
illustrate the problem.

The results below are obtained using JikesRVM version 3.1.2~\cite{JikesRVMWeb}.
While our benchmarks can run on any JVM, we use JikesRVM for these experiments
both because it can report many detailed measurements, and because it allows
direct comparison with our new algorithm, which we implemented in JikesRVM.
Section~\ref{sec:results} contains a detailed description of the experimental
setup and methodology.

\subsection{Existing cache implementations}

Google's Guava library is a widely-used infrastructure for implementing software
caches. It implements a simple get/put interface for keys and values, and offers
a variety of eviction policies to manage the capacity of the cache. Even with
this library, however, there are several significant challenges to obtaining
good cache performance:

\vspace{0.3em}
\noindent
\textbf{Choosing a cache size is difficult.} The most straightforward eviction
trigger is capacity-based: the programmer chooses the maximum number of entries
(key-value pairs) that the cache will hold. When the cache grows beyond this
limit, it evicts entries in least-recently-used order.

The challenge of this policy is how to choose a good size: too small and the
cache will underperform; too large and the program will slow significantly or
crash due to memory pressure. In many cases, the cached values vary widely in
size, so the same set of entries could account for wildly varying quantities
of data.

One potential solution is to measure representative workloads during testing and
configure the cache accordingly (e.g., by assuming an average size key and
value). Unfortunately, this approach is brittle: unless the workload is
extremely uniform and predictable, the number of entries is not a reliable
predictor of the memory footprint of the cache. If actual workloads in
deployment differ substantially, then performance will suffer.

\vspace{0.3em}
\noindent
\textbf{There is no easy way to measure memory footprint.} To handle these
cases, Guava can manage cache capacity in terms of an application-specific
``weight''. The programmer implements a \texttt{weigh()} method that can compute
a weight value for any entry. The weight method could, for example, count the
number elements in a container. Entries are evicted in LRU order to keep the
total weight under the limit. Example code is shown in
Figure~\ref{fig:guava-code}.

\begin{figure}
\begin{small}
\begin{lstlisting}
Cache<Key, Graph> graphs = CacheBuilder.newBuilder()
  .maximumWeight(100000)
  .weigher(new Weigher<Key, Graph>() {
     public int weigh(Key k, Graph g) {
        return g.vertices().size();
     } 
   } ).build(
        new CacheLoader<Key, Graph>() {
           public Graph load(Key key) {
             return createExpensiveGraph(key);
          } } );
\end{lstlisting}
\end{small}
\caption{Guava cache that stores graphs and uses a weighing function to
  represent their size.}
\label{fig:guava-code}
\end{figure}

Implementing an accurate weighing method, however, is not always easy. Ideally,
we would like to know the exact size (in bytes) of each cached value. In the
case of simple structures, such as strings, an accurate size is easy to
compute. Measuring the size of complex data structures is more difficult. One
problem is encapsulation: it might not be possible to access the hidden
implementation of a class. Another problem is structure sharing: when measured
independently, the shared substructures could be counted multiple times,
distorting the total weight.

One alternative is to explicitly measure the size of a data structure at runtime
using reflection. This is the approach taken by JAMM, which is based on the JVM
Tool Interface~\cite{jamm}. While accurate, its cost is so high that it is not
practical for use in production settings. For example, measuring a data
structure with 1 million objects can take 5 seconds of wall clock time, and causes the benchmark to run
over 100$\times$ slower than the approach we propose here.

\vspace{0.3em}
\noindent
\textbf{Eviction doesn't work.} The purpose of eviction is to control cache
memory use by freeing low-value entries. In a garbage collected language,
however, eviction does not achieve this goal. The cache can remove entries and
null out all references to them, but the memory is not actually reclaimed until
a garbage collection occurs. Guava attempts to address this problem by
performing eviction lazily, but we have observed cases where eviction actually
makes memory pressure \emph{worse}. If the cache misses on a recently-evicted
entry, then it will create a new one, resulting in two copies in memory at the
same time. A secondary effect, which we show at the end of this section, is an
increase in the allocation rate. One of our observations is that it \emph{only}
makes sense to do eviction at collection time, when memory is actually reclaimed.

\vspace{0.3em}
\noindent
\textbf{Soft references don't do the right thing.} Recognizing that caches can
be a source of memory pressure, Guava also offers the option of storing entries
in soft references. The Guava documentation claims that soft references are
reclaimed in LRU order when memory is tight. While this policy is not required
by the Java standard, we found that Oracle's HotSpot JVM does implement such a
strategy~\cite{oraclesoftrefs}.\cut{ in a clear attempt to help caching
  application.} It has two serious limitations, however: first, the policy is
hard-wired to LRU, and second, the LRU ordering is global for all soft
references. Large Java programs, such as web applications, can have multiple
caches, and the global LRU order is particularly problematic if these caches are
accessed with different frequencies. Entries in a less-frequently used cache all
wind up at the end of the LRU queue, resulting in the entire cache being
dumped. Figure~\ref{fig:cache-freq}(a) illustrated this effect: the bigger the
differences between the caches, the worse the impact of the soft reference
policy.

\subsection{Exploring cache-GC interaction}

\noindent
To study these problems in detail, we implemented a simple key-value store in
Java that we can drive with a range of workloads and under a range of
conditions. Our goal is to isolate cache performance and its interaction
with the garbage collector. In the context of a larger application, these effects
might be hard to separate from unrelated program behavior.

\vspace{0.3em}
\noindent
\textbf{Trace files.} The input to the driver is a trace file
that specifies the workload as a sequence of key-value requests. The keys are
just names, but the values represent data structures of varying sizes. The goal
is to model caching of data that has a non-trivial structure (as opposed to
strings, for example). Real-world examples might include a parsed XML document
in tree form, or memoized computations in an optimizing compiler. The trace file
itself just specifies the size of each tree in number of nodes.
Figure~\ref{fig:trace} shows an example fragment of a trace file. 

We generate each trace file according to a set of parameters: (1) number of
unique keys, (2) minimum and maximum sizes of the values, (3) the distribution
of value sizes, (4) the number of key requests (trace length), and (5) the
temporal distribution of keys in the trace. 

The set of unique key/value pairs is generated by choosing value sizes at random
from a Pareto distribution. Many kinds of workloads, including web requests and
file accesses, have been found to follow this kind of power law
distribution~\cite{atikoglu2012, cunha95, newman05}.  The sequence of key
requests is also drawn from a Pareto distribution, which governs the temporal
locality of the trace. Here, higher alpha values create more locality, and lower
values spread out the distribution more evenly. We use an alpha value of 0.1,
which is on the low end and requires caches to be larger to achieve a high hit
rate. The traces range in length from 10,000 to 50,000 key requests, with 2000
to 5000 unique keys.

\begin{figure}
\centering
\begin{small}
\begin{BVerbatim}
key_8        179074
key_12       180434
key_1        150999
key_188      126021
key_2        154588
key_28       119220
...
\end{BVerbatim}
\end{small}
\caption{Example trace file. The number associated with each key determines
the \textbf{size} of the data structure that is stored as the value.}
\label{fig:trace}
\end{figure}

\vspace{0.3em}
\noindent
\textbf{Execution.}  Our driver initializes the cache, reads the trace and sends
the cache a sequence of get and put operations. Its behavior is configured using
several parameters: (1) size of the cache (number of entries for Guava), (2) the
max heap size, and (3) the cost of a miss. The miss cost models the time to
fetch data from a remote source or recompute it, which is proportional to the
size of the resulting data. In these experiments the miss cost only represents
the time it would take to transmit the data over a 10MB/s network
connection. Higher miss costs (for example, modeling a more expensive
computation or a database query) would only exaggerate the shape of our
graphs. The trace is processed as follows:

\begin{enumerate}

\item For each key in the trace, call \texttt{Cache.get(K)}.

\item If it \textbf{misses}, the driver uses the value number in the trace file
  to construct a tree of the given size. It delays execution for a time
  proportional to the size of the tree and the miss cost. It then stores the key
  and value (tree) in the cache using \texttt{Cache.put(K,V)}.

\item If it \textbf{hits}, the cache returns the associated tree data
  structure. The driver performs a modest computation on the tree that visits
  all the nodes.

\end{enumerate}

\vspace{0.3em}
\noindent
\textbf{Measurements.}
We record several measurements for each complete run of a trace file:

\begin{itemize}

\item Total time, mutator time, and GC time (with GC time broken down into sub-categories).

\item Hit and miss rate, as well as time spent servicing misses

\item Number of garbage collections

\item Total memory allocation

\end{itemize}

\subsection{Guava performance}

\begin{figure}
\centering
{\small \textbf{(a)} Workload of small values (10K-50K)}
\includegraphics[width=\columnwidth]{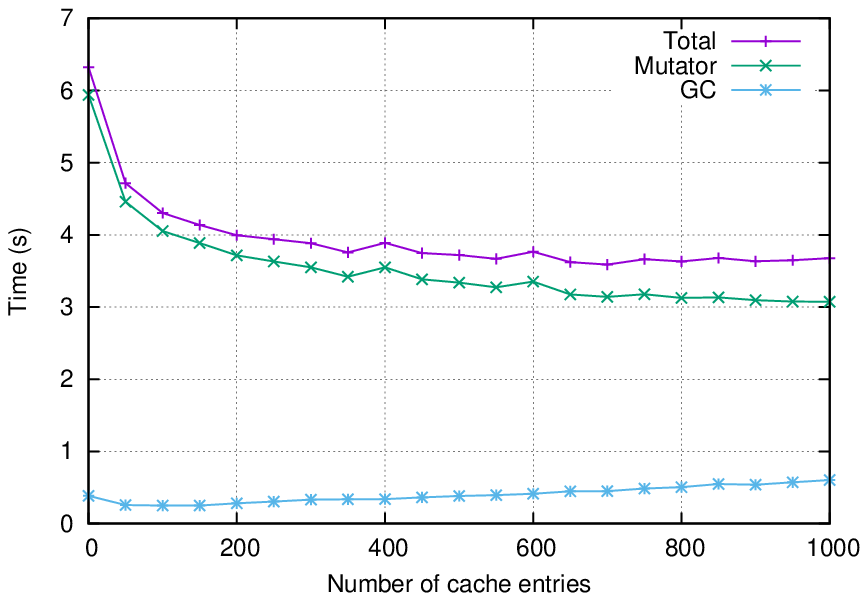}\\
\vspace{1.0em}
\centering
{\small \textbf{(b)} Workload of medium values (50K-100K)}
\includegraphics[width=\columnwidth]{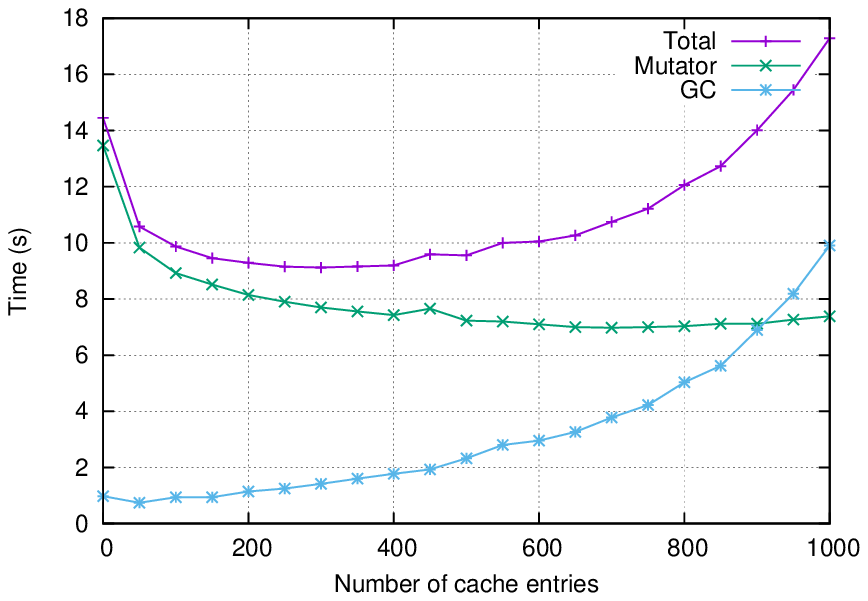}\\
\vspace{1.0em}
\centering
{\small \textbf{(c)} Workload of large values (100K-200K)}
\includegraphics[width=\columnwidth]{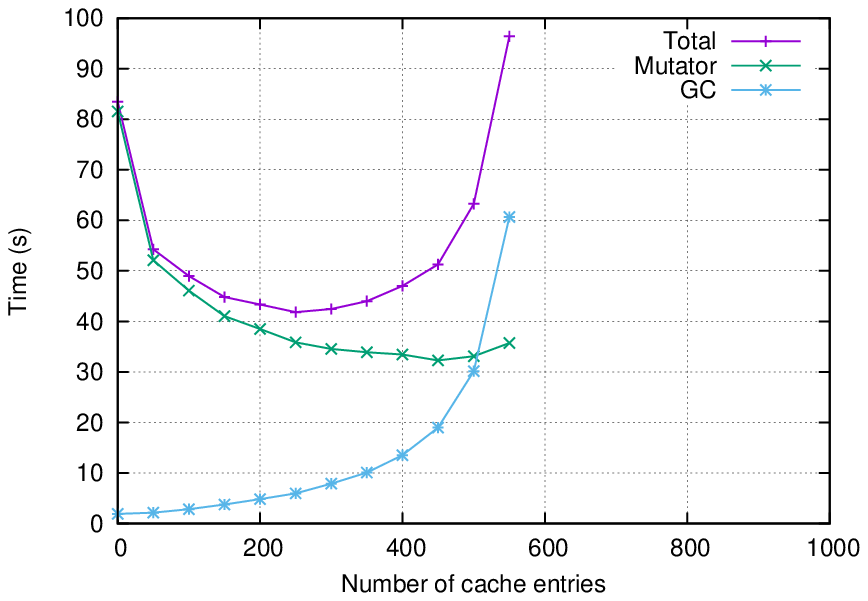}\\
\caption{Guava performance under three workloads: choosing a
  good number of entries is difficult.}
\label{fig:guava-time}
\end{figure}

The three graphs in Figure~\ref{fig:guava-time} are typical of the performance
we see for a traditional cache implementation. For these experiments, we vary the capacity of
the cache from 100 to 1000 entries (the X axis) and measure the time to process
the entire trace (Y axis). The heap size is fixed at 115MB. We show mutator
(application) time, GC time, and total time. The three graphs differ in the sizes
(in bytes) of the cached values, which affects both the memory footprint of the
cache and the cost of a cache miss:

\begin{itemize} 

\item Figure~\ref{fig:guava-time}(a) shows the performance on a trace with
  small-sized values (10K to 50K bytes). In this case, the total time continues
  to drop all the way out to 1000 entries, suggesting that the cache could
  probably accommodate more before incurring a memory cost.

\item Figure~\ref{fig:guava-time}(b) shows the same graph for medium-sized value
  (50K to 100K bytes). It exhibits the typical ``bowl'' shape for the total time,
  which is explained by the opposing curves of the miss cost (going down) and GC
  time (going up). Miss costs are accounted for in the mutator time. GC costs at
  the right edge go up steeply because the cache is approaching the maximum heap
  size and causing memory pressure.

\item Figure~\ref{fig:guava-time}(c) shows the results for larger values (100K
  to 200K bytes). Under this workload, only a narrow range of cache sizes is
  usable. Too few and the miss costs are huge; too large and the cache runs out
  of memory.

\end{itemize}

\begin{figure}[h]
  \includegraphics[width=\columnwidth]{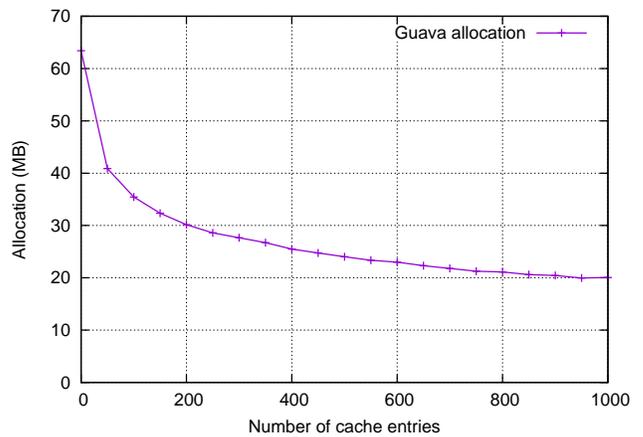}
  \caption{Undersizing a cache (left side) incurs the cost of more misses
    as well as the cost of increased allocation.}
  \label{fig:guava-stats}
\end{figure}

\noindent
Looking at other metrics provides further insight into this behavior. At the
left of the graph (when the number of entries is small) several factors are
hurting performance. First, the number of misses is higher, incurring the cost
of ``fetching'' (rebuilding) the value. Second, evictions are more frequent,
filling up memory with garbage. Third, rebuilding the values increases total
allocation costs. Figure~\ref{fig:guava-stats} plots the total amount of
allocation for a run of the trace under different cache sizes. The smaller sizes
cause a significantly higher allocation rate.

These results suggest that bounding a cache unnecessarily when more memory is
available can lead to performance degradations that are nearly as great as
having a cache that is too large. In fact, we conclude that it does not make
sense to evict entries at all if there is sufficient memory to hold them.

\section{Prioritized Garbage Collection}
\label{sec:prioritized}

In this section, we introduce new runtime support for software caches and other
space-sensitive data structures. The primary goal is to provide an effective
mechanism for implementing eviction policies that take into account both memory
utilization (which only the garbage collector knows) and the relative value of cache
entries (which only the application knows).

\subsection{API}

Prioritized garbage collection is a cooperative technique, so the central
feature of our system is an API that allows application code to communicate
directly with the garbage collector. Our API is modeled after Java
\emph{reference objects}, which already play a similar role. Each reference object
points to a single target referent, and the particular type of reference object
chosen tells the collector how to treat the referent -- typically, it
specifies when the referent can be reclaimed even if it is still reachable. For
example, the Java garbage collector will clear a \texttt{WeakReference} to an
object when there are no other ordinary (strong) references to it. Since the
application is not aware of when collection occurs, it discovers that a
reference has been cleared only when it attempts to get the referent and the
result is null.

\subsubsection{PrioReference}

\begin{figure}[ht]
\begin{lstlisting}
class PrioReference<T> extends 
      java.lang.ref.Reference<T>
{
  // -- Constructor: the new reference belongs 
  //    to the given space
  PrioReference(T obj, PrioSpace<? super T> space);

  // -- Get the referent
  public T get(); 

  // -- Get and set the priority
  public void setPriority(int new_prio);
  public int getPriority();

  // -- Inquire about the memory footprint
  public boolean hasGCSize();
  public int getGCSize();
}
\end{lstlisting}
\caption{A priority reference holds a single referent with a given priority. The
  application can also inquire about the total amount of memory reachable
  through this reference.}
\label{fig:priorityreference}
\end{figure}
 
Following this model, a \emph{priority reference} is a new reference type
that may be cleared by the collector in order to bound memory use, but only
after \emph{all other references of lower priority have already been
  cleared}.  The \texttt{PrioReference} class definition is shown in
Figure~\ref{fig:priorityreference}.  Each \texttt{PrioReference} has a
priority value, which is simply an integer -- higher values represent
higher priorities.  The application is free to choose and change this value
in any way. In our cache implementation (below), for example, each cached
value is held in a \texttt{PrioReference}, and we implement LRU eviction by
ensuring that the most recently hit entry has the highest priority.

The \texttt{getGCSize()} method allows the application to find out the
total memory footprint of all objects reachable \emph{only} through this
reference. Ordinarily, this information is difficult or inefficient to
compute, but our collector computes it as part of its normal marking
phase, incurring little overhead. It needs to know this information in
order to determine whether the size bound has been exceeded, so we opt to
make it available to the application as well. The only caveat is that it is
only computed at GC time, and so it is not guaranteed to be fresh or even
computed at all. The \texttt{hasGCSize()} method asks whether or not the
collector has computed the size. Once \texttt{getGCSize()} is called, the
flag is reset until the next collection.

In Section~\ref{ssec:collection}, we describe the details of how size information is
computed and used in the collection algorithm.

\subsubsection{PrioSpace}

\begin{figure}[!b]
\begin{lstlisting}
class PrioSpace<T>
{
  // -- Create a priority space with a specific 
  //    target size in bytes
  PrioSpace(int bound);

  // -- Create an adaptive size priority space
  //    expressed in terms of percentage of the heap, 
  //    either as fraction used or a fraction free.
  PrioSpace(float fraction, boolean used_or_free);

  // -- Get the actual memory footprint of the whole 
  //    space after GC
  public boolean hasGCSize();
  public int getGCSize();
}
\end{lstlisting}
\caption{A priority space holds a set of priority references and governs their
  lifetime collectively under single policy.}
\label{fig:priorityspace}
\end{figure}
 
In order to support multiple independent caches, priority references
are grouped into \emph{priority spaces}, each with its own memory
bound. The \texttt{PrioSpace} class definition is shown in
Figure~\ref{fig:priorityspace}. \texttt{PrioSpaces} are not spaces in
the memory management sense, but rather a collection of references that are
considered together. The collector considers each priority space
separately, evicting the lowest priority references until the total
memory footprint reachable from the remaining references is smaller
than the target bound. At the lowest level, this bound is expressed in
bytes, but our API also allows it to specified as a fraction of total
memory or as a fraction of available memory, which can change
dynamically as the program runs.

In a typical configuration, each cache would be managed in its own
\texttt{PrioSpace}, but this one-to-one mapping is not required. For example, we
can emulate SoftReferences by placing all \texttt{PrioReferences} in one
\texttt{PrioSpace} and associating the priority of a reference with the time the
program last uses a reference.

\subsection{Measuring Memory Footprint}

\noindent
The job of our collector is to bound the total memory footprint of each priority
space by keeping as many high-priority entries as will fit in the available
space, and freeing the rest. In order to do this job the collector must be able to
accurately measure the memory footprint of each entry, as well as its
contribution to the total memory footprint of the priority space. There are
several factors that complicate this computation. First, entries may consist of
complex, pointer-based data structures, so it is not sufficient to measure only
the size of the object directly pointed to by the priority reference. Second, we
need to properly handle shared structures to avoid counting them multiple
times. Third, we need to account for fragmentation to make sure that the sum of
the sizes properly reflects the actual fraction of total memory used.

\subsubsection{Fragmentation}

\noindent
Many kinds of memory allocators can suffer from \emph{fragmentation}, in which
small chunks of memory become unusable, taking away from the total
available. Fragmentation is a concern for our algorithm because it could cause
us to underestimate the total memory cost of a set of
objects. Traditionally, fragmentation is divided into two categories:
\emph{internal} and \emph{external}~\cite{wilson95}. Our algorithm can easily
account for internal fragmentation, but external fragmentation is more complex,
as discussed below. In general, though, fragmentation has not been found to be a
major problem for dynamic memory management~\cite{johnstone98}.

Internal fragmentation is created when the allocator reserves more memory than
is requested for an object in order to comply with alignment, padding, or size
restrictions. For example, our current implementation uses a free list allocator
with fixed size classes, so all small objects must be allocated into one of the
51 possible size denominations. Objects that do no fit perfectly are allocated
in the next size up, leaving some number of bytes unused. Luckily, internal
fragmentation is easy to account for: whenever the algorithm needs the size of
an object, denoted $size(o)$ in this section, we use the \emph{allocated} size,
which includes unused padding, if any.

External fragmentation is created when sequences of allocations and
deallocations leave unused memory in between objects. Unlike internal
fragmentation, our algorithm cannot directly account for external fragmentation
because it cannot determine if a given unused fragment should be charged to any
particular priority space (or none at all). As a result, our algorithm views all
free space as available for non-cache objects, potentially reducing the usable
free space by the amount of the fragmentation. In practice, the free list
allocator in MMTk has been observed to have low
fragmentation~\cite{blackburn04}, but if it became a problem we could change the
underlying allocator or collection algorithm, both of which are largely
orthogonal to our technique. There is no reason we could not use a compacting
collector, for example.

\subsubsection{Reachability}
\noindent
If we consider only the memory used by the objects immediately pointed to by
each priority reference, computing size is easy and requires little additional
collector machinery. However, many data structures have complex internal
structure. Even strings typically consist of two objects: a string
object and a character array. From a memory use standpoint, it makes sense for
the size of a string to include the size of its character array.

We therefore define the total memory footprint of a priority space as the sum
(in bytes) of the sizes of all objects transitively reachable \emph{only} from
the priority references in that space. We purposely exclude any objects that are
also reachable from roots because clearing the associated priority references
will not cause them to become garbage. In other words, we exclude the parts of
the memory footprint that the priority space cannot control.

\subsubsection{Structure Sharing}
 
\noindent
The size computation has algebraic properties that are crucial to ensuring we
meet the target memory bound. If the size computed for a particular referent
\texttt{o} is $size(o)$, the total size of a set of objects $S$ should be given
by the following formula:

$$size(S) = \sum_{i = 0}^{|S|}{size(o_i)}$$

\noindent
While the formula seems obvious, consider the case where two priority references
share some internal structure. We need to be careful that common objects are
only counted once. Otherwise, the sum of the sizes could overestimate the total
memory footprint. Our algorithm guarantees this property by visiting each object
only one time (see algorithm below for details), but this choice affects the
measured sizes of the individual entries. For example, if two structures with
roots $o1$ and $o2$ share a common object $p$, only one of their sizes will
account for $p$ -- the one that is traced first. In this case, $o_2$ will not
consider any objects reachable from $p$. Let $o\backslash p$ represent
the objects only reachable from $o$. Then the sizes of $o_1$ and $o_2$ are as follows:

$$size(o_1) = size(o_1\backslash p) + size(p)$$
$$size(o_2) = size(o_2\backslash p)$$

The sum, however, still accurately reflects how much memory is actually in
use by all of the priority references together:

$$size(o_1) + size(o_2) = size(o_1\backslash p) + size(p) + size(o_2\backslash p)$$

\noindent
The property above is crucial to our enforcement mechanism because it means we
can trace a sequence of cache entries in any order, and the running sum
of their sizes at any point represents how much memory would be occupied if all
remaining entries were evicted.

To see why tracing the sequence in order is important, consider an
extreme example in which three small entries, $o_1$, $o_2$, and $o_3$,
each of size $K$ bytes alone, share a large common structure of size
$L$ bytes. Visiting the references in order will cause $o_1$ to have
size $L+K$, and $o_2$ and $o_3$ to have size $K$. If the space bound
is less than $L$ it might be tempting to evict only $o_1$. This choice
will not achieve the expected space savings because $o_2$ and $o_3$
hold references to the shared state so only $K$ bytes will be
recovered. If we inspect them in order, though, we can see that the
bound is reached during tracing of $o_1$, and all three entries need
to be evicted in order to get below target size $L$.

A similar issue exists for structures shared between priority spaces, although
we consider this case to be more unusual.  At each collection, the
\texttt{PrioSpace} learns the total memory footprint of its
\texttt{PrioReferences}. Consider the case where two \texttt{PrioSpaces} hold
\texttt{PrioReferences} to the same data $o$ with memory footprint of $K$. bytes
At first, we may consider that the total memory footprint of both
\texttt{PrioSpaces} includes those $K$ bytes. However, we use the marking bit to
measure $o$, implying we can only measure $o$ once per collection. While each
collection processes all \texttt{PrioSpaces}, only the first \texttt{PrioSpace}
to measure those $K$ bytes and add it to their total memory footprint. Any other
space will regard $o$ as a structure in use elsewhere in the program and pass
over it.

\subsection{Collection Algorithm}
\label{ssec:collection}
\noindent
The Prioritized Garbage Collector is built on a standard full-heap
mark/sweep collector. The algorithms are amenable, however, to any
\emph{tracing} collector, including copying collectors and generational
collectors. The reason that we focus on pure mark/sweep is that caches are
highly non-generational data structures: none of the entries are
short-lived, and under LRU, entries must sit in the cache for some
time before they become the least-recently used and are evicted. In
addition, this collector performs full-heap collections more frequently, so
eviction policies built on it run more frequently.

Our collection algorithm is based on two key ideas:

\begin{itemize}

\item We reorder heap tracing so that we can visit specific regions of the heap
  graph based on their reachability -- specifically, the regions reachable from
  priority references. No significant additional work is required, so this
  overhead is very low. This technique has been used by other systems to check
  heap properties using the garbage collector~\cite{aft09}.

\item We introduce \emph{bounded marking}, in which the garbage collector traces
  a region until a condition is met (for example, the memory bound is reached);
  then it simply stops marking and nulls out potential dangling
  pointers. Unmarked objects are reclaimed immediately by the sweeper without
  being touched again.

\end{itemize}

\noindent
The key to our algorithm is that it does not explicitly free low priority
references to satisfy the memory bound; rather, it \emph{protects} high priority
references (by marking them) until the memory footprint grows to the bound. Once
the bound is reached, the algorithm ceases to mark any other objects in the
priority space. With a small amount of fixup to avoid dangling pointers, the
remaining low priority references will be reclaimed immediately by the
sweeper. This approach \emph{guarantees} that the memory bound will be respected
(see reasoning below). A secondary benefit is that the collector does not need
to touch the evicted objects, which might improve CPU cache performance,
although we do not measure this effect here.

To ensure that the collector counts only objects reachable solely from the
priority space, we reorder the phases of the collector as follows:

\begin{enumerate}[label=Phase (\arabic*):,leftmargin=*]

\item Compute exact target sizes (in bytes) for priority spaces that
  are specified in fractional terms. For example, if a priority space
  specifies its bound as 20\% of the total heap, the collector uses
  computes $0.2 * heapsize$ as the bound.
  
\item Premark all \texttt{PrioReference} objects held by the
  \texttt{PrioSpace}.

\item Perform a complete mark over transitive closure from the root
  references. This process will stop at each \texttt{PrioReference}, so the
  only objects that will remain unmarked are either garbage or are
  reachable \emph{only} through a priority reference.

\item Revisit each \texttt{PrioReference} in priority order, from
  highest to lowest, and perform a transitive closure starting at its
  referent object. During this phase, the collector computes a running sum
  of the object sizes. If the total size hits the memory bound
  for the space then this phase ends immediately. Since it visits the
  priority references in order, the remaining unmarked instances must all
  be lower priority.

\item Any \texttt{PrioReferences} with unmarked referents are nulled
  out (the pointer to the referent is set to null).

\item If Phase (3) ended early, it could leave \emph{part} of a data
  structure marked, with outgoing pointers to unmarked objects. To preserve
  memory safety the collector also nulls out all potentially dangling pointers
  as well as the priority reference containing this partial structure -- in
  effect, evicting the entire structure.

\item Evicted objects are garbage, and can be immediately reclaimed by
  the sweeper.

\end{enumerate}

\vspace{0.5em}

\noindent
Notice that since the collector is doing the work, eviction only occurs at GC
time. But this makes sense: we cannot assess the global memory situation
until GC time, and we don't have an effective way to recycle memory
in between GCs.

\paragraph{Partial eviction.} Phase (5) of the algorithm handles the case in
which bounded marking stops part-way through a data structure, leaving pointers
from marked objects to unmarked objects. In order to preserve memory safety,
this phase nulls out all of these references. In addition, we null out the
\texttt{PrioReference} itself, which makes even the partially marked portion of
the structure unreachable.

There is a case, however, in which a program could observe a partially evicted
structure. We believe this case would be rare, however, since it only happens
under very specific conditions. If the program creates a weak reference to an
object in the cache, and that object is part of the marked portion of the
partially evicted structure, then our JVM will preserve the weak reference and
the object it points to. This weak reference will be cleared at the next GC,
since the strong reference from the cache entry is now unreachable, but there is
a brief period where the program could follow the weak reference and find a
structure with null fields in unexpected places.

We have not been able to find a satisfactory and performant solution to this
problem. If a program cannot tolerate partial eviction, however, our
implementation provides a switch that forces eviction only at cache entry
boundaries -- that is, when the space limit is reached, the mark phase continues
until the current entry is completely marked. The downside of this option is
that it allows the cache to grow beyond the size limit. In practice, however, we
find that it only hurts performance when cache entries are very large (100's of
KB each) and keeping the extra objects is a substantial burden on memory
resources.

\subsection{Overheads}

\subsubsection{Runtime Overhead}

\noindent
Phase (3) of the algorithm above requires the collector to visit
\texttt{PrioReference} objects in priority order.  In our current
implementation, the \texttt{PrioSpace} class keeps its priority references
in a max heap, so that common operations are $O(logN)$ time. This cost,
however, is paid every time the priority of a reference changes: the
priority space must re-insert it into the heap. It is possible that in some
configurations the total cost of these inserts would exceed the cost of
simply sorting the list immediately before garbage collection. In practice,
we have not observed a significant performance penalty.

\subsubsection{Space Overhead}

\noindent
Each \texttt{PrioSpace} stores its \texttt{PrioReferences} in a max heap, which
is implemented as an array.  Furthermore, we arrange the \texttt{PrioSpaces}
into a linked list in the VM, using an extra reference in each PrioSpace to
point to the next one. So, if we have $N$ \texttt{PrioReferences} spread across
$K$ \texttt{PrioSpaces}, then the total space overhead is $N + K$ references, in
addition to the space required for each \texttt{PrioReference} and
\texttt{PrioSpace} instance.

\subsection{Sache: A Space-aware Cache}

\noindent
Using these mechanisms, we implemented a space-aware cache that we call a
\emph{Sache}. The interface to the Sache is essentially the same as the Guava
LRU cache, and it can be used as a drop-in replacement. An
overview of the Sache class is shown in Figure~\ref{fig:sache}.

\begin{figure}[t]
\begin{lstlisting}
class Sache <K,V> extends HashMap<K, PrioReference<V>> {

  protected long highest_priority;
  protected PrioSpace<V> priospace;

  public Sache(long maxSize);

  public boolean put(K key, V value);
  public V get(K key);
  public V remove(K key);

  private void update();
}
\end{lstlisting}
\caption{Interface for \texttt{Sache} space-aware cache}\label{fig:sache}
\end{figure}

As with a cache built around a HashMap, we can \texttt{put} key-value
pairs in, \texttt{get} a value using a key, and \texttt{remove} a
value using a key. The Sache stores all values in \texttt{PrioReference}
objects, so that the collector can measure and evict them as necessary.
The Sache increments the \texttt{highest\_priority} value as necessary to ensure
that the most recently hit value has the highest priority. The methods work
as follows:

\vspace{0.3em}
\noindent
\textbf{Constructor}: create an empty hash map and an empty
\texttt{PrioSpace} with the given space bound \texttt{maxSize}. We also
support a version that specifies the size as a fraction of available memory.

\vspace{0.3em}
\noindent
\textbf{get(K)}: If there is an entry in the map for the given key,
update the priority on its \texttt{PrioReference} to \texttt{highest\_priority}
(incrementing if necessary) and return the referent value. If not, return
null.

\vspace{0.3em}
\noindent
\textbf{put(K, V)}: Create a new \texttt{PrioReference} in the Sache's
priority space with the value V as the referent and give it the highest
priority. Store the pair of key K and priority reference in the hash map.

\vspace{0.3em}
\noindent
\textbf{remove(K)}:
If there is an entry in the map for the given key, remove it from the map and
remove the corresponding \texttt{PrioReference}.

\vspace{0.3em}
\noindent
\textbf{update}: 
Periodically scan the hash map looking for entries that have null values in
their \texttt{PrioReferences}, indicating that they were evicted by the
garbage collector. This method runs when the program accesses
the Sache after a collection and does not interact with the \texttt{PrioSpace}.
Since the collector already evicted the \texttt{PrioReferences} from the
\texttt{PrioSpace}, we can complete this operation in $O(N)$, where $N$
is the number of entries in the Sache.

It is possible to add collector support to remove an entry from the Sache
whenever the collector decides to evict its corresponding \texttt{PrioReference}.
In particular, we can manipulate the Sache's outgoing pointers directly.
This would improve the performance of the \texttt{update} operation.
However, we decided against this to provide a more general reference type versus
a cache-specific reference type.

\subsection{Adaptive Sizing}

\noindent
The Sache is an efficient and effective bounded-size cache supported directly by
the garbage collector. The problem remains, however, of how to choose its
size. As we show in Section~\ref{sec:results}, choosing a fixed size, measured as
a fraction of total bytes in the heap, yields good performance across a range of
workloads for our simple key-value store. Many applications that use caching,
however, are not just key-values stores -- they have other computations going on
that are competing for resources. For example, looking at the results we might
choose a Sache sized to occupy about half of the heap. If other parts of the
program need the other half, however, the resulting memory pressure will cause
massive GC overhead.

Our solution is to adaptively size the Sache according to available memory. Our
current policy is simple: at each garbage collection we choose a Sache size
that ensures a minimum amount of free memory (if possible). Other policies are
certainly possible, and we discuss some of them in the future work section.

To ensure a free memory reserve of size $R$ bytes, we need to know the size of the
heap ($H$ bytes), and the total live size of all the data \emph{not} in the Sache
($L$). Using this information, the maximum size of the Sache is set to $H - (L+R)$.

We can efficiently recompute this value at every garbage collection, growing or
shrinking the Sache adaptively. Phase 2 of our algorithm visits all objects not
in the Sache, so we can augment it to also compute their total size, which is
$L$. Before Phase 3 starts, we compute the target bound for the Sache, and the
algebraic properties of the bounding mechanism guarantee that evicted entries will
leave at least R bytes free.

\section{Results}
\label{sec:results}

In this section, we evaluate how  prioritized garbage collection helps
overcome the conflicting space-time tradeoffs of software caching and
automatic memory management. Note that there is nothing particularly
innovative about our core caching algorithm or data structure. The Sache is
essentially just a hash map. The difference is in how it manages
provisioning -- specifically, how it manages the number of entries in the
hash map. Given a \emph{fixed} configuration, it performs almost
identically to the Google Guava cache on a given workload. What we show
here is that with support from the collector, the Sache can
automatically \emph{vary} its configuration to make the best use of
available memory regardless of the characteristics of the workload. In
addition, it can adapt online to changes in memory use or workload.

\subsection{Experiments}

We evaluate our system using two caching applications: a key-value store
driven by the same synthetic workloads presented in
Section~\ref{sec:problem}, and a web caching implementation driven by real
traces of web traffic. Since the Sache API is almost identical to the Guava
LRU cache API, we can easily switch between cache implementations in each
benchmark system. We show results for the following experiments:

\begin{itemize}

\item First, we repeat the experiments shown in Section~\ref{sec:problem}
  using a Sache configured with a range of fixed sizes
  (non-adaptive). Specifying the size in terms of memory footprint instead of
  number of entries is robust across a range of workloads.

\item Second, we enable the adaptive sizing algorithm for the Sache and show
  that it can dynamically shrink or grow the Sache in response to changing
  memory demands in other parts of the program.

\item Third, we test the two cache implementations under a real workload of web
  traffic. Software caches perform the work of the central data structure in a
  web cache (something similar to \texttt{memcached}). This application is challenging to
  implement in a garbage-collected language, since the goal is to cache as much
  as possible.

\end{itemize}

\subsection{Methodology}

We implemented our GC mechanisms in JikesRVM version 3.1.2 as a modification to
the stop-the-world mark/sweep collector. The Sache itself is implemented at the
application level. We have also incorporated these mechanisms into a
generational collector, but it provides little benefit for cache-heavy
applications because caches are so strongly non-generational. In addition,
the generational collector delays computation of the size information and
enforcement of the eviction policies, since these mechanisms cannot be
implemented properly for partial collections.

\paragraph{Building and running.}
We build JikesRVM in the FastAdaptive configuration (for performance). The Guava
experiments are run on an unmodified version of JikesRVM to avoid incurring any
possible penalty related to prioritized GC. All experiments are run on a
machine with dual 2.8GHz Xeon X5660 processors (X64) with a total of 12GB of
main memory running Ubuntu Linux kernel 3.2.0.

We run each configuration only one time, but due to the length of the traces
(tens of thousands) small fluctuations in the cost of any individual cache
operation or garbage collection are averaged out. In addition, the performance
differences we highlight are orders of magnitude greater than experimental noise,
and in many cases it is the difference between running to completion or crashing
with an out-of-memory error.

\subsection{Non-adaptive Sache}

\begin{figure}
\centering
{\small \textbf{(a)} Workload of small values (10K-50K)}
\includegraphics[width=\columnwidth]{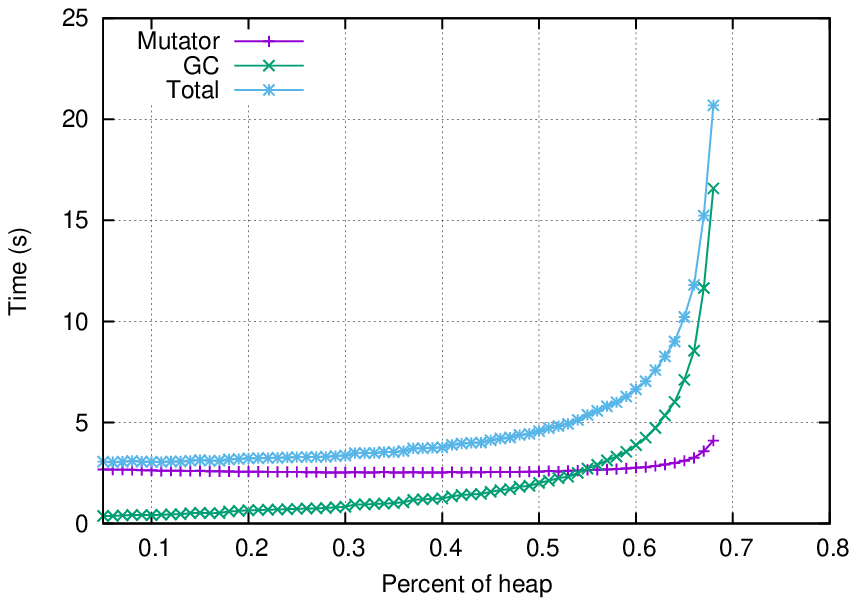}\\
\vspace{1.0em}
\centering
{\small \textbf{(b)} Workload of medium values (50K-100K)}
\includegraphics[width=\columnwidth]{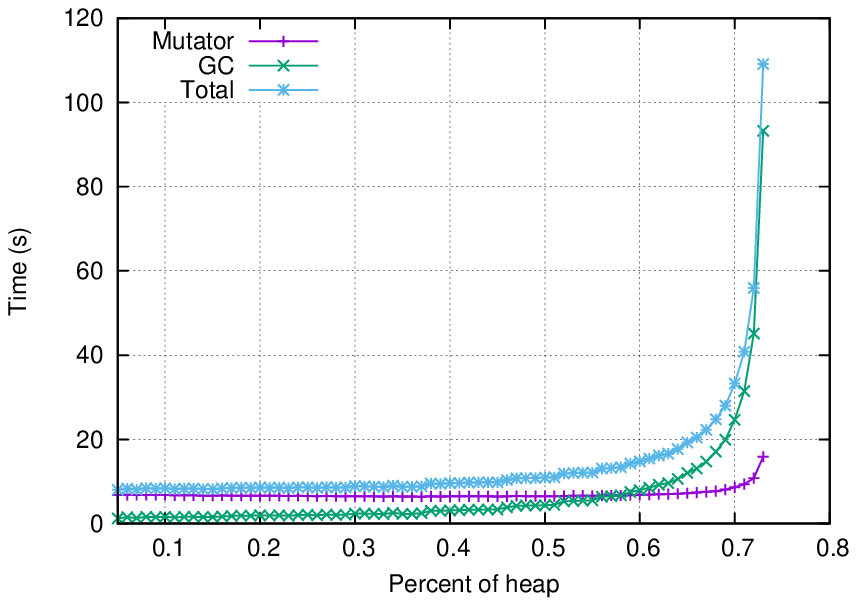}\\
\vspace{1.0em}
\centering
{\small \textbf{(c)} Workload of large values (100K-200K)}
\includegraphics[width=\columnwidth]{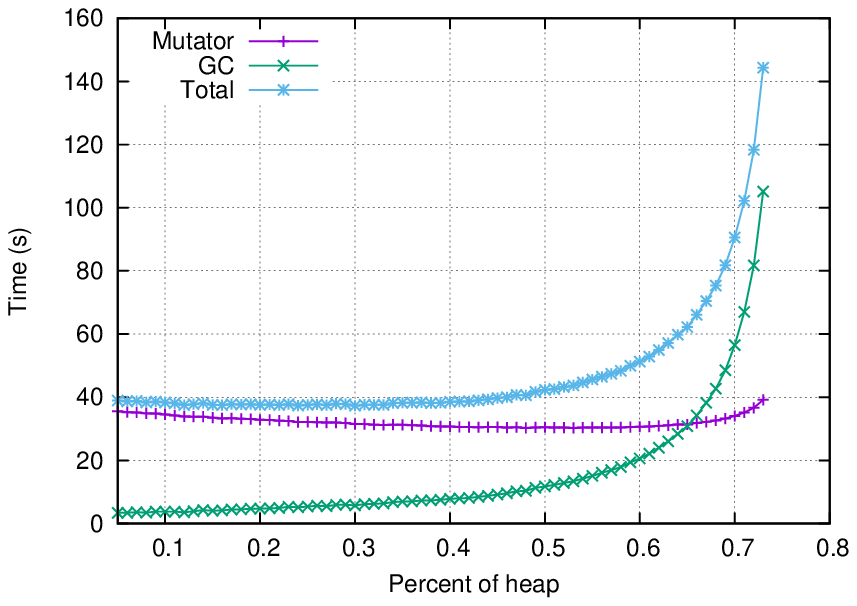}\\
\caption{Sache performance under three different workloads. While absolute
  performance varies, compared to the Guava cache the space-time tradeoff is
  relatively independent of the workload. The Sache makes the best use of the
  available space.}
\label{fig:sache-time}
\end{figure}

Figure~\ref{fig:sache-time} shows the performance of a Sache under the same workloads
as the Guava cache. We measure the size as a percentage of the heap instead of a maximum
number of entries. Each graph has three curves: one for mutator time, one for GC time, 
and another for total time.

\begin{itemize}

\item Figure~\ref{fig:sache-time}(a) shows the performance on the trace with small-sized
  values. The total time is a bowl shape, but the minimum occurs early on
  at a limit 10\% of the heap size. Afterwards, the mutator time remains steady
  until 65\%.

\item Figure~\ref{fig:sache-time}(b) shows the performance on the trace with
  medium-sized values. The minimum of the bowl is not present on the graph. Just
  like the graph with small-sized values, the Sache peaks heavily towards the
  end, dominated by garbage collection cost.

\item Figure~\ref{fig:sache-time}(c) shows the same graph for larger values. The Sache
 continues to function even when using 75\% of the heap. This comes at the cost
 of increased garbage collections.

\end{itemize}

All three graphs have the same shape. The mutator time is about the same until after 60\% 
of the heap is reserved for the Sache. Recall that the Sache limit is only enforced at a 
collection. Therefore, the Sache holds more values and more hits occur. The application does
not rebuild values, lowering the mutator cost. On the other hand, the Guava caches have to 
rebuild many items at when we have a low limit on the number of entries. 

Despite the Sache using a lot of memory prior to GC, the GC curve starts low and peaks 
towards the end. The prioritized GC frees elements of the Sache as soon as it observes
the limit would be exceeded. This leaves much less garbage in the heap than simply evicting
the value from the Sache outside of GC. 

\begin{figure}[h]
  \includegraphics[width=\columnwidth]{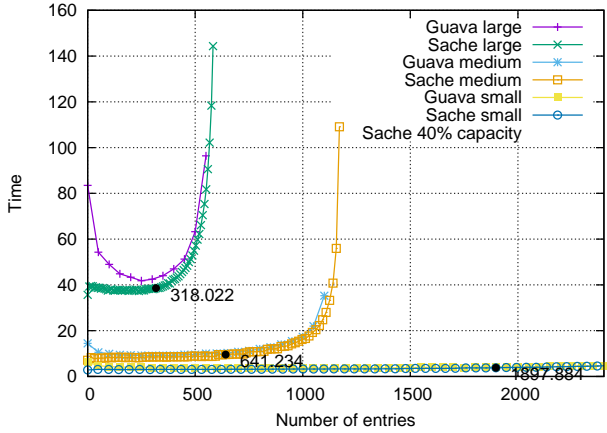}
  \caption{Comparing Sache and Guava LRU on three workloads: the
    performance is very similar, but the three Sache curves represent the
    same configuration choices. The highlighted points represent a Sache set
    to 40\% of the heap, which easily accommodates all three workloads by
    using different numbers of entries.}
  \label{fig:guava-sashe}
\end{figure}

Figure~\ref{fig:guava-sashe} compares the above results with the graphs for
Guava in Section~\ref{sec:problem}. To directly compare them, we plot
the Sache with the average number of entries after a garbage collection. Recall
that prioritized GC only enforces the bound at a collection, so the Sache
actually grows larger in between collections. These numbers are approximately
the memory bound of the Sache divided by the average size of the values in the
trace.

We see that the shapes of the curves between Guava and the Sache are about
the same. The key difference is that Sache curves represent the \emph{same
  set of configuration options} -- 10\% of the heap on the low end and 80\%
of the heap on the high end. A programmer could choose a Sache capacity of
40\% and be able to achieve good performance regardless of the
workload. The three labeled points on the graph show how this choice leads
to different numbers of entries under the different workloads.

Looking at the small workload, the Guava's curve is about flat while the Sache's extends
further out and eventually starts to move upward. The Sache utilizes as much memory as it
can before the limit is enforced. This allows the lower end limits of 5\% to perform better
than the hard limit of 1500 entries on Guava.

Looking at the larger values, the Guava cache fits less than 600. The size limit on the
Sache and prioritized GC allow the Sache to handle a mix of sizes. In particular, the
collector frees values that the Sache will not keep because of the size limit. This also
allows the Sache to hold more items than the Guava cache can, by prioritizing smaller
values.

\subsection{Multiple Caches}

\begin{figure}
  \centering
  \begin{subfigure}{.5\textwidth}
    \caption{Guava Caches with Soft References}
    \includegraphics[width=0.9\columnwidth]{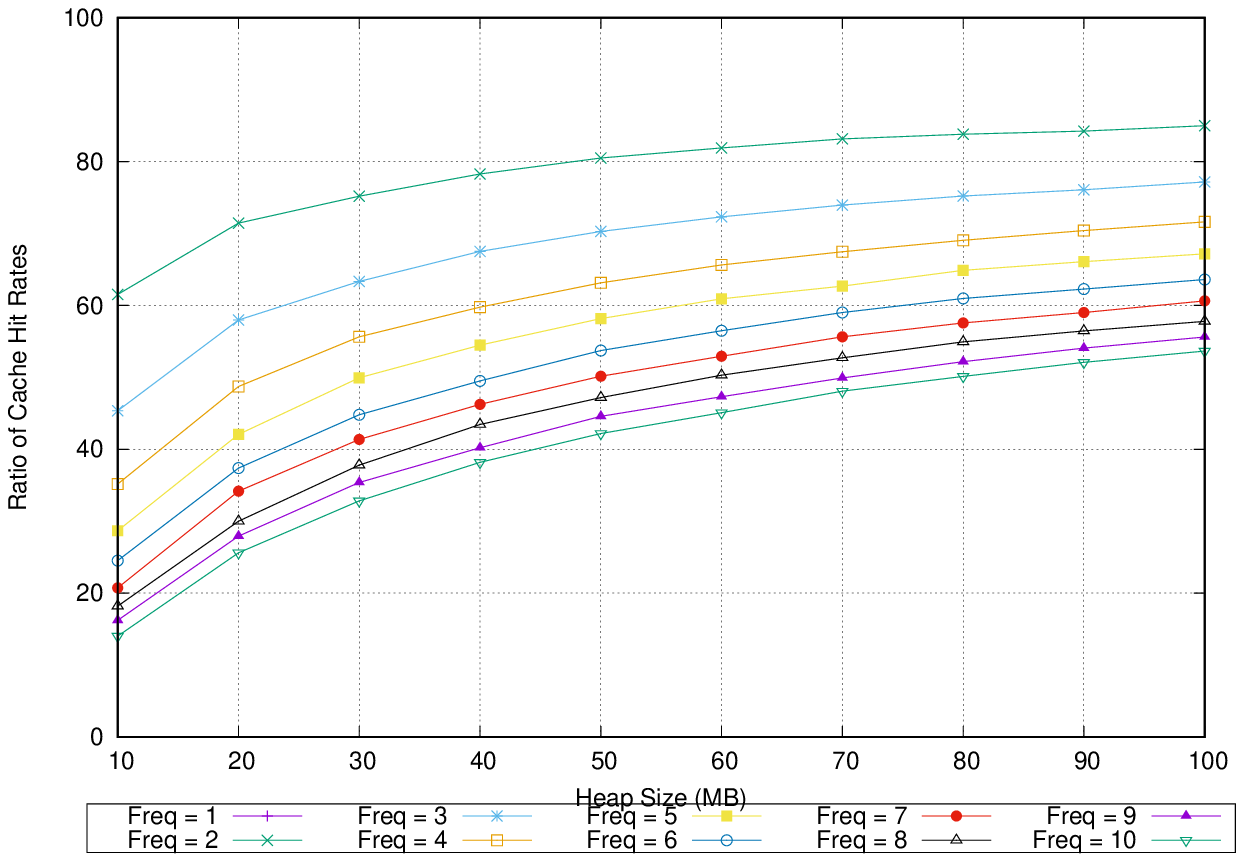}
    \label{fig:soft-freq}
  \end{subfigure}
\begin{subfigure}{.5\textwidth}
  \caption{Saches with PrioReferences}
  \includegraphics[width=0.9\columnwidth]{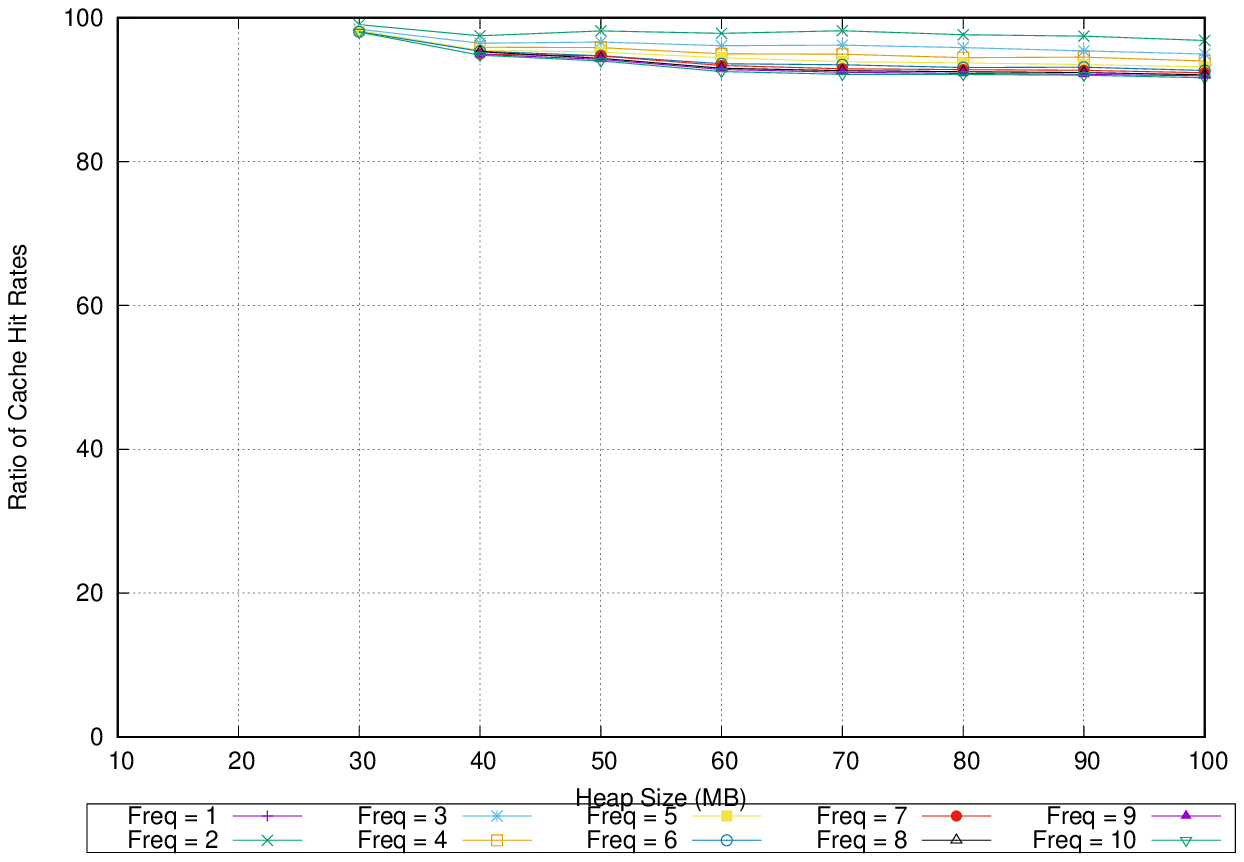}
      \label{fig:prio-freq}
\end{subfigure}
\begin{subfigure}{.5\textwidth}
  \caption{Maximum Hit Rate of the Caches}
  \includegraphics[width=0.9\columnwidth]{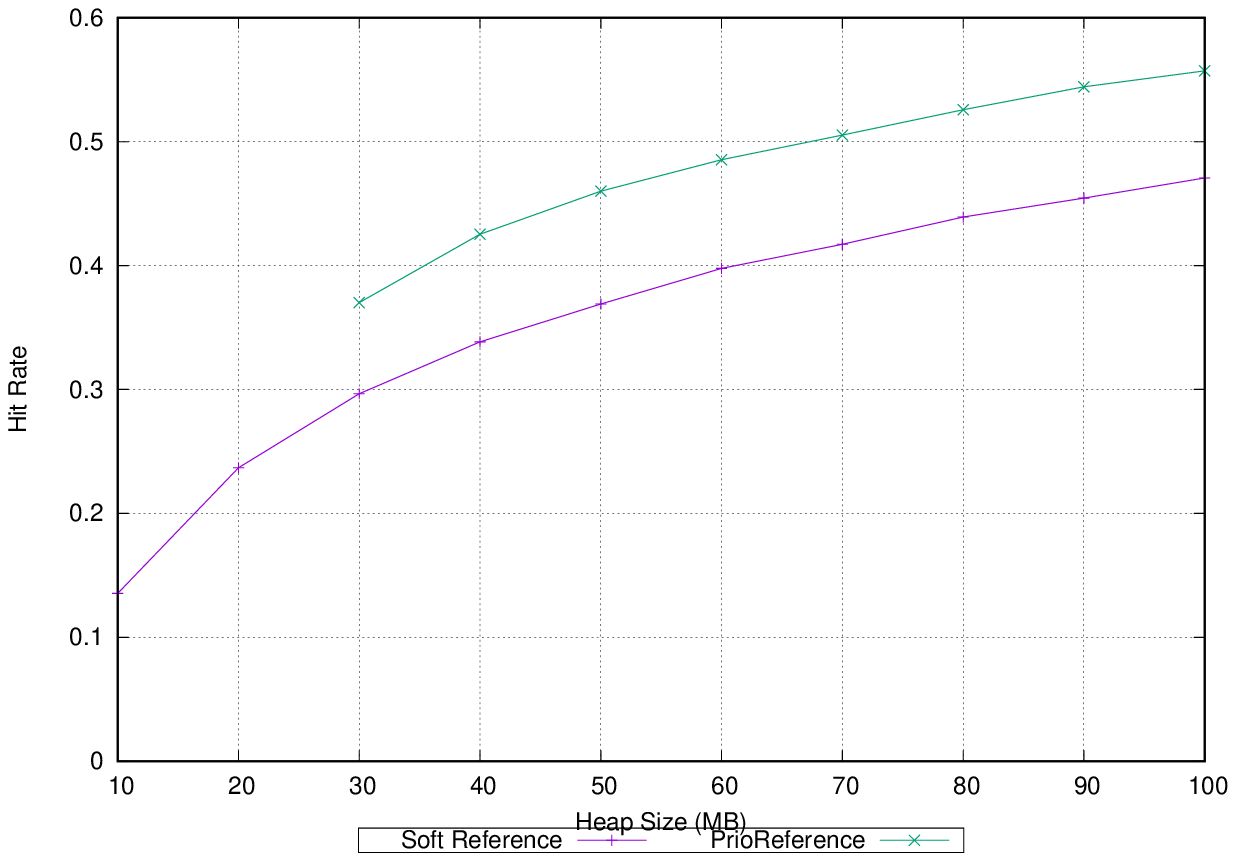}
      \label{fig:max}
\end{subfigure}
\caption{Hit rates can drop dramatically when soft references are used for two
  caches working at different frequencies. Prioritized garbage collection keeps
  the hit rates of both caches relatively close by managing their resources
  separately.}
\label{fig:cache-freq}
\end{figure}

One of the problems with soft references is that they are managed by the JVM
using a single global policy. Even if that policy happens to be the right one
(e.g., LRU) treating all soft references as equal can lead to very bad
performance. Consider, for example, a program with two caches. If one cache is
accessed less often than the other then its entries will tend to appear towards
the end of the global soft reference LRU queue. When memory is tight, many more
of its entries will be reclaimed regardless of their value to the application
(i.e., regardless of the miss cost).

We measure this effect directly using the following experiment: we run two
caches simultaneously and have both serve requests from our largest trace, but
at different frequencies. One cache processes $N$ requests for every request the
other cache processes. We also use a larger heap size to measure the effect of
having more memory available. We run this experiment on the Hotspot VM with
Guava caches using soft references and no explicit size limit or eviction
policy. We do this to measure the effectiveness of HotSpot's soft reference
eviction policy alone.  We also run the experiment on our modified VM with two
Saches, each configured to use 20\% of the heap. We use the LRU policy for each
Sache's \texttt{PrioSpace} to match Hotspot's policy for removing soft
references. Hit rates are reported as a number between 0 and 1. We report the
difference between hit rates on the same scale. Figure~\ref{fig:cache-freq}
presents the results of this experiment.

Figure~\ref{fig:cache-freq}\subref{fig:soft-freq} shows the results for soft
references running on HotSpot, which clearly reflect the global LRU policy. The
X-axis shows the heap size; the Y-axis shows the ratio of the measured hit rate
to the maximum hit rate. As the difference between the access frequencies of the
two caches grows, the hit rate of the slower cache drops significantly. Its
entries appear to be less valuable because they are less frequently accessed, so
the soft reference eviction policy removes them first. The effect is more
pronounced in smaller heaps because the soft reference policy is more
aggressive. In the worst case (10-to-1 frequency difference), the hit rate is
only 1/4 of its potential, but the degradation at just a 2-to-1 difference is
very significant as well.

Figure~\ref{fig:cache-freq}\subref{fig:prio-freq} shows the same results for
prioritized garbage collection. The hit rates of the two Saches differ by at
most 5\% because the PrioSpaces manage their references separately, so the VM
does not clear the PrioReferences in the less frequently used cache regardless
of what is going on in the higher frequency cache.

Finally, Figure~\ref{fig:cache-freq}\subref{fig:max} shows the maximum hit rate
(in absolute terms) for both the Guava cache with soft references and the Sache
with PrioReferences. As expected, with no competing memory demands, the two
systems perform almost identically. Note that for
Figures~\ref{fig:cache-freq}\subref{fig:prio-freq} and \ref{fig:cache-freq}\subref{fig:max}, the Sache does not have data for 10MB and
20MB. Since VM objects share heap space with Java applications in JikesRVM, we
needed 30MB to start running the experiment without running out of memory.

\subsection{Adaptive Sache}

The purpose of the adaptive sizing algorithm is to allow the cache to respond to
changes in the available resources. Our goal is to prevent the cache from
competing with other application data structures, causing unnecessary memory
pressure.

For these experiments, we modified our benchmark to build a separate large data
structure that grows as the trace is processed. Each experiment is divided into
three phases: during the first 1/3 of the trace, no extra memory is used; during
the middle 1/3, the program starts growing the non-cache data structure,
consuming more and more memory; during the last 1/3 of the trace, the program
slowly dismantles the structure, allowing the collector to alleviate the pressure.

We ran our medium-sized-objects trace through the key-value store using both the
Sache and Guava LRU cache. For the Sache, the adaptive algorithm is configured to
target a 50\% memory reserve. This value corresponds to a heap two times the
live size, which is a good target for performance~\cite{1094836}. We size the
Guava cache using the data collected in Figure~\ref{fig:guava-time}(b): the best
size for this workload appears to be around 350 entries. We fix the heap at
115MB, as in the earlier experiments.

\begin{figure}
\centering
\begin{minipage}{0.47\textwidth}
\centering
\includegraphics[width=\textwidth]{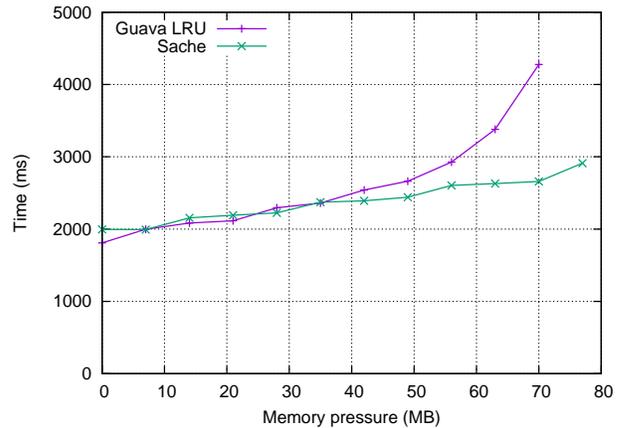}
{\small (a) Total time}
\end{minipage}\hfill
\begin{minipage}{0.47\textwidth}
\centering
\includegraphics[width=\textwidth]{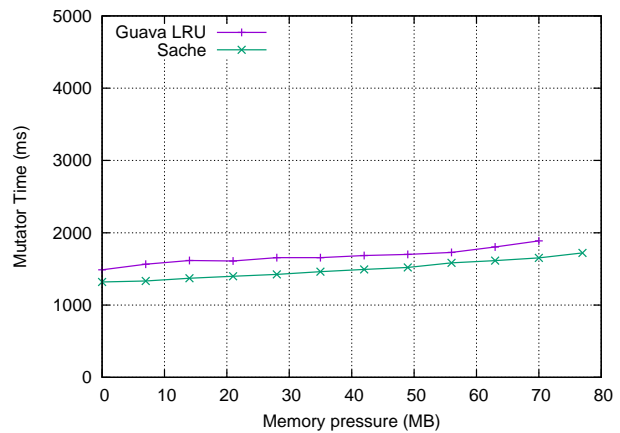}
{\small (b) Mutator time}
\end{minipage}
\centering
\includegraphics[width=0.47\textwidth]{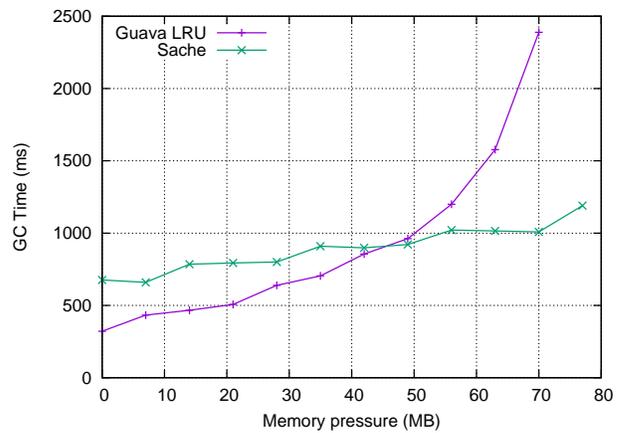}
{\small (c) GC time}
\caption{Performance of Sache vs Guava LRU cache under increasing memory
  pressure: our adaptive sizing algorithm shrinks the Sache to avoid triggering
  massive GC overhead. At 77MB, the application with the Guava LRU cache crashes.}
\label{fig:pressure}
\end{figure}

Figure~\ref{fig:pressure} shows the total time, mutator time, and GC
time. Unsurprisingly, the Sache and the Guava cache exhibit similar performance
as long as memory is plentiful. As memory pressure increases,
however, the Guava cache competes with non-cache structures in memory, and GC
costs skyrocket. When memory pressure exceeds 77MB, the Guava implementation
crashes. The Sache automatically shrinks to ensure sufficient free memory,
resulting in a smooth curve and no crashes.

\begin{figure}
\includegraphics[width=\columnwidth]{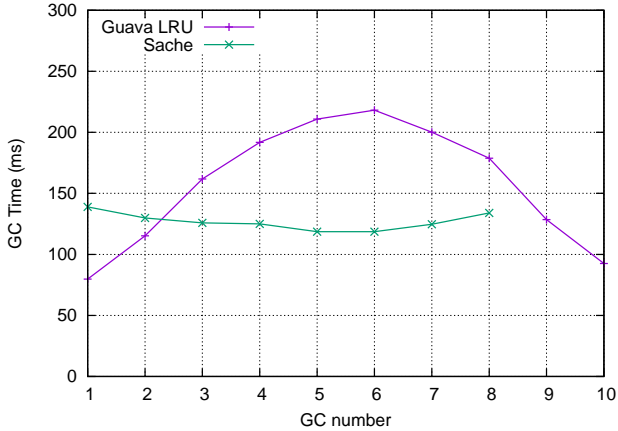}
\caption{GC time over a single run: without the ability to adapt, cache and
  non-cache structure compete, triggering costly GCs.}
\label{fig:onerun}
\end{figure}

Figure~\ref{fig:onerun} provides some insight into this behavior. It shows the
GC time for each collection during a single run of the benchmark. As expected,
in the Guava implementation, once the non-cache structure begins to grow, each GC
becomes much more expensive. In addition, memory scarcity triggers more frequent
GCs. With the adaptive Sache, the GC time is flat: the algorithm guarantees that
the cache will not cause the live size to exceed the target reserve.

\subsection{Web Caching Workload}

\begin{figure}
\includegraphics[width=\columnwidth]{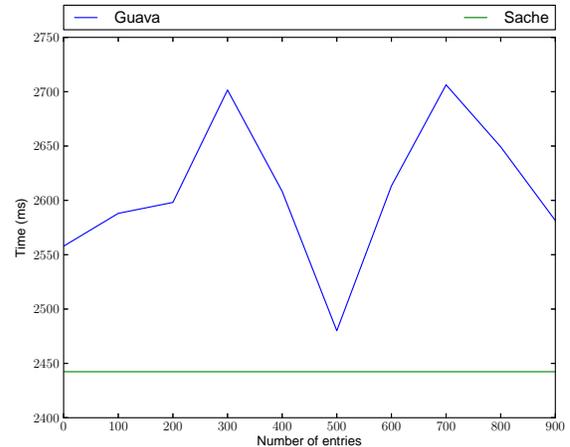}
\caption{Performance of the Sache and Guava cache on real web traffic traces
  across a range of cache sizes.}
\label{fig:webtraffic}
\end{figure}

The heart of a web caching application, such as \texttt{memcached}, is a key-value cache
like the ones we describe here. \cut{We could not find an implementation of memcached
in Java, and we believe that the reason is that garbage collection supports
caching so poorly.} We adopted the techniques for testing web caches and applied
them to our cache implementations. We use the BU 272 trace, a record of real web
traffic, to drive the caches, and measure performance as above\cite{cunha95}. It consists of
15K entries requesting a total of 72MB of web data. Figure~\ref{fig:webtraffic}
shows the performance of the Guava cache across a range of numbers of
entries. The Sache is a flat line, since it chooses its own size.

\section{Related Work}
\label{sec:related}

\subsection{Reference Types}
Hayes introduced Ephemerons to determine unreachable objects in key-property lists instead of
using a list with a weak key and strong value~\cite{hayes97}. This capability allows entries
with dead keys to be removed from the list and properly deleted. We expect items in a
cache to be reachable only from the cache. Therefore, removing stale objects is not a
priority. Instead, we remove elements that cause our structure to exceed its limit.

\subsection{Programs Acting on Resource Limits}
There has been considerable work on caching for web traffic. The Greedy-Dual
algorithm takes into account the amount of time to obtain a page as well as the
size of the page to determine its eviction~\cite{cao97}. The cached objects are
text documents, so their size can be easily measured. Our system allows these
kinds of eviction policies to be used in general software caches, where measuring 
the size of a cached structure is non-trivial.

Yang and Mazi\`{e}res describe resource containers for Haskell that bound the
memory usage of untrusted code ~\cite{yang14}. Exceeding the limit of the
container kills the accompanying thread. A Sache is a more general structure
which allows the program to choose what action to take when memory bounds are
exceeded. In principle, it could also be used to implement a similar security policy.

Czajkowski and von Eicken introduced an interface for programmers to monitor the
resources used by threads in a Java program~\cite{czajkowski98}. Furthermore,
they allow programmers to implement their own reaction to threads exceeding
their resource limits. They use bytecode rewriting to track heap memory usage
for each thread. In contrast, we track a list of known objects and use the
garbage collector to both track memory usage and enforce the limits. In
addition, our technique uses general heap reachability to define bounded
structures, rather than thread ownership.

JAMM uses JVMTI to traverse the heap and compute the size of data
structures~\cite{jamm}.  This approach is flexible and powerful, but very
slow. A large data structure can require seconds of runtime to size (according
to the documentation). By piggybacking on the garbage collector, we can perform
the same measurement with almost no performance overhead. The tradeoff, however,
is that we cannot compute sizes at arbitrary points during execution.

Price, Rudys, and Wallach divide a process into tasks and used the garbage
collector to track how much memory is attributed to each
task~\cite{price03}. Our work expands on this by allowing arbitrary data
structures to be tracked and by providing a way to enforce a size limit.

\cut{
GC Assertions allow programmers to efficiently check predicates about the
structure of the heap by using the garbage collector~\cite{aft09}. Our work
expands on this idea by allowing the program to interact with the garbage
collector in richer ways: asking quantitative questions, such as size, and
asking the collector to enforce policies.
}

\subsection{Using GC to Assist Running Programs}

O'Neill and Burton presented simplifiers as a way to improve the performance
of a program ~\cite{o'neill06simp}. Objects can add a \lstinline{simplify()} 
method that the garbage collector invokes when the collector traces over it.
The Data Structure Aware Garbage Collector lets the program denote which objects
are internal nodes for data structures~\cite{cohen15dsagc}. It uses this information 
to improve garbage collection for these structures and therefore overall performance.
Our work uses the GC to allow programs to run when they would run out of memory in normal
execution. Furthermore, we can traverse and modify the structures without editing
those structures' code.

\section{Conclusions and Future Work}
\label{sec:conclusion}

This paper presents a new approach to managing the conflicting tradeoffs between
software caching and garbage collection. The key to our approach is widening the
interface between the application and the runtime system in order to increase
cooperation and break the tradeoffs. Caches are representative of a broader
class of resource-sensitive data structures and algorithms that are not well
served by existing collection algorithms. Our ongoing work is to look carefully
at the particular needs of these applications and provide ways for them to be
more effectively served by the runtime system to improve performance and
robustness. We also are looking into how to divide available heap space amongst
multiple PrioSpaces and therefore multiple Saches.

\acks
Emery Berger was supported by NSF grant CCF-1439008. Diogenes Nunez was supported
by the Google Research Award.

\bibliographystyle{plain}

\begin{thebibliography}{}

\end{thebibliography}


\begin{thebibliography}{21}

\bibitem{aft09}
Edward~E. Aftandilian and Samuel~Z. Guyer.
\newblock {GC} assertions: Using the garbage collector to check heap
  properties.
\newblock In {\em Proceedings of the 2009 ACM SIGPLAN Conference on Programming
  Language Design and Implementation}, pages 235--244. ACM, 2009.

\bibitem{atikoglu2012}
Berk Atikoglu, Yuehai Xu, Eitan Frachtenberg, Song Jiang, and Mike Paleczny.
\newblock Workload analysis of a large-scale key-value store.
\newblock In {\em Proceedings of the 12th ACM SIGMETRICS/PERFORMANCE Joint
  International Conference on Measurement and Modeling of Computer Systems},
  SIGMETRICS '12, pages 53--64, 2012.

\bibitem{jamm}
Jonathan Bellis.
\newblock Jamm.
\newblock https://github.com/jbellis/jamm.

\bibitem{blackburn04}
Stephen~M. Blackburn, Perry Cheng, and Kathryn~S. McKinley.
\newblock Myths and realities: the performance impact of garbage collection.
\newblock In {\em Proceedings of the International Conference on Measurements
  and Modeling of Computer Systems}, pages 25--36, 2004.

\bibitem{cao97}
Pei Cao and Sandy Irani.
\newblock Cost-aware www proxy caching algorithms.
\newblock In {\em Proceedings of the USENIX Symposium on Internet Technologies
  and Systems on USENIX Symposium on Internet Technologies and Systems}, pages
  18--18, 1997.

\bibitem{cohen15dsagc}
Nachshon Cohen and Erez Petrank.
\newblock Data structure aware garbage collector.
\newblock In {\em Proceedings of the 2015 ACM SIGPLAN International Symposium
  on Memory Management}, ISMM 2015, pages 28--40, New York, NY, USA, 2015. ACM.

\bibitem{cunha95}
Carlos Cunha, Azer Bestavros, and Mark Crovella.
\newblock Characteristics of {WWW} client-based traces.
\newblock Technical Report BU-CS-95-010, Computer Science Department, Boston
  University, Boston, MA, USA, 1995.

\bibitem{czajkowski98}
Grzegorz Czajkowski and Thorsten von Eicken.
\newblock {JRes}: A resource accounting interface for {Java}.
\newblock In {\em Proceedings of the 13th ACM SIGPLAN Conference on
  Object-oriented Programming, Systems, Languages, and Applications}, OOPSLA
  '98, pages 21--35, 1998.

\bibitem{hayes97}
Barry Hayes.
\newblock Ephemerons: A new finalization mechanism.
\newblock {\em SIGPLAN Not.}, 32(10):176--183, October 1997.

\bibitem{1094836}
Matthew Hertz and Emery~D. Berger.
\newblock Quantifying the performance of garbage collection vs. explicit memory
  management.
\newblock In {\em OOPSLA '05: Proceedings of the 20th annual ACM SIGPLAN
  conference on Object oriented programming, systems, languages, and
  applications}, pages 313--326, New York, NY, USA, 2005. ACM.

\bibitem{androidsoftrefs}
Google Inc.
\newblock {\em SoftReference | Android Developers}, 2016 (Accessed March 23,
  2016).

\bibitem{JikesRVMWeb}
{Jikes RVM}.
\newblock {IBM}, 2005.
\newblock http://\-jikesrvm.\-sourceforge.\-net.

\bibitem{johnstone98}
Mark~S. Johnstone and Paul~R. Wilson.
\newblock The memory fragmentation problem: Solved?
\newblock In {\em Proceedings of the 1st International Symposium on Memory
  Management}, pages 26--36, 1998.

\bibitem{mitchell03leakbot}
Nick Mitchell and Gary Sevitsky.
\newblock Leakbot: An automated and lightweight tool for diagnosing memory
  leaks in large {Java} applications.
\newblock In {\em {ECOOP} 2003 - Object-Oriented Programming, 17th European
  Conference, Darmstadt, Germany, July 21-25, 2003, Proceedings}, pages
  351--377, 2003.

\bibitem{newman05}
M.~E.~J. Newman.
\newblock Power laws, {Pareto} distributions and {Zipf's} law.
\newblock {\em Contemporary Physics}, 2005.

\bibitem{o'neill06simp}
Melissa~E. O'Neill and F.~Warren Burton.
\newblock Smarter garbage collection with simplifiers.
\newblock In {\em Proceedings of the 2006 Workshop on Memory System Performance
  and Correctness}, MSPC '06, pages 19--30, New York, NY, USA, 2006. ACM.

\bibitem{oraclesoftrefs}
Oracle.
\newblock {\em SoftReference ({Java} Platform {SE} 6)}, 2015.

\bibitem{price03}
David~W. Price, Algis Rudys, and Dan~S. Wallach.
\newblock Garbage collector memory accounting in language-based systems.
\newblock In {\em Proceedings of the 2003 IEEE Symposium on Security and
  Privacy}, SP '03, pages 263--, Washington, DC, USA, 2003. IEEE Computer
  Society.

\bibitem{wilson95}
Paul~R. Wilson, Mark~S. Johnstone, Michael Neely, and David Boles.
\newblock Dynamic storage allocation: {A} survey and critical review.
\newblock In {\em International Workshop on Memory Management}, pages 1--116,
  September 1995.

\bibitem{xu11leakchaser}
Guoqing Xu, Michael~D. Bond, Feng Qin, and Atanas Rountev.
\newblock {LeakChaser}: Helping programmers narrow down causes of memory leaks.
\newblock In {\em Proceedings of the 32nd ACM SIGPLAN Conference on Programming
  Language Design and Implementation}, PLDI '11, pages 270--282, 2011.

\bibitem{yang14}
Edward~Z. Yang and David Mazi\`{e}res.
\newblock Dynamic space limits for {Haskell}.
\newblock In {\em Proceedings of the 35th ACM SIGPLAN Conference on Programming
  Language Design and Implementation}, PLDI '14, pages 588--598, New York, NY,
  USA, 2014. ACM.

\end{thebibliography}





\end{document}